# Enhanced magnetoelectric coupling in two-dimensional hybrid multiferroic heterostructures

Xilong Xu,[1] Li Yang*,[1,2]

[1]Department of Physics, Washington University in St. Louis, St. Louis, Missouri 63130, USA

[2]Institute of Materials Science and Engineering, Washington University in St. Louis, St. Louis, Missouri 63130, USA

**Abstract**

Magnetoelectric coupling in insulating multiferroic materials is invaluable for both fundamental research and multifunctional device applications. However, material realization remains a significant challenge. We employ first-principles calculations to predict an enhanced and stacking-robust magnetoelectric coupling in a multiferroic van der Waals heterostructure consisting of $CrCl_3$ and $CuCrP_2S_6$. Within this heterostructure, the intralayer ferroic orders exhibit characteristics of a type-I multiferroic, while the interlayer ferroic orders resemble those of a type-II multiferroic, thus creating a hybrid-type multiferroic. Notably, the interlayer magnetic configuration can be switched upon electric polarization reversal, while maintaining the insulating property without any nontrivial phase transitions. Our analysis reveals a strongly interfacial spin local-field effect and proximity polarization effect on band alignments. These factors work synergistically to decide the interlayer exchange interaction and drive enhanced magnetoelectric coupling. Finally, we show that this magnetoelectric effect is robust in multi-layer structures, exhibiting a dependence on the layer number's parity. This research offers promising prospects for discovering enhanced magnetoelectric couplings in hybrid multiferroic van der Waals heterostructures.

*E-mail: lyang@physics.wustl.edu

The magnetoelectric (ME) effect, allowing for the effective manipulation of magnetism through an electric field and vice versa in multiferroic materials, holds immense promise in energy-efficient devices, spanning information storage, spintronic devices, memory, and magnetic sensors [1-6]. Traditionally, multiferroics are categorized into two types according to their ME couplings [7,8]. In Type-I multiferroics, where ferromagnetism (FM) and ferroelectricity (FE) have independent origins, ME coupling tends to be weak [9,10]. On the other hand, Type-II multiferroics exhibit strong ME coupling, as FE ordering consistently coincides with a magnetic transition [11-15]. However, achieving this significant coupling often involves challenges such as the stringent material specifications, weak FE polarization, and relatively low working temperature [16,17]. Despite extensive research efforts, identifying robust multiferroic materials with enhanced ME coupling remains a considerable challenge.

Recently, van der Waals (vdW) multiferroic heterostructures have garnered increasing research interest [18-28]. For example, the magnetic anisotropic energy (MAE) of $CrGeTe_3$ can be tuned by attaching FE $In_2Se_3$ [29]. In bilayer $CrI_3$, the interlayer magnetic order can be adjusted by attaching FE $Sc_2CO_2$ or $In_2Se_3$, where the FE component serves as a gating field to tune the interlayer band alignment of magnets [30-33]. However, these predictions either rely on charge-transfer effects, leading to emerging metallic states and unfavored leakage current, or are based on calculations of specific interlayer supercell structures due to lattice mismatch and considerable strain. More recently, intrinsic ME coupling via FE and antiferroelectricity (AFE) orders has been proposed, but switching between FE and AFE orders remains challenging [34].

In this study, we predict that the lattice-matched $CrCl_3$/$CuCrP_2S_6$ (CCPS) heterostructure exhibits enhanced insulating ME coupling. The interlayer magnetic configuration efficiently switches between FM and antiferromagnetic (AFM) couplings by altering FE polarization of CCPS. This effect remains robust across different interlayer stacking styles of the widely observed monoclinic phase. We elucidate two mechanisms: a unique spin local-field effect, where the switch in electric polarization significantly alters the spin component of interfacial chalcogen atoms and interlayer superexchange paths, and a substantial variation in band alignment induced by proximity electric polarization, modifying the hopping energy of interlayer superexchange interactions. These two factors synergistically enhance the overall ME effect. Finally, beyond bilayer, we predict a layer-dependent ME effect in multilayer heterostructures.

The first-principles simulations are performed using the Vienna ab initio Simulation Package (VASP) [35]. The atomic structure optimization is performed within the generalized gradient approximation (GGA) and Perdew–Burke–Ernzerhof (PBE) exchange-correlation functional [36,37] until the force on each atom is less than 0.02 eV/Å. To better capturing the correlation effects and underestimated band gaps by density functional theory (DFT), the electronic structure calculations are calculated by the hybrid functional theory (HSE06) [38,39]. Meanwhile, DFT+U calculations have been done, and the main results are not changed within the reasonable range of U values [40,41]. The k-mesh is set to 7 x 7 x 1 with a 450-eV cutoff energy. Dipole moment correction is employed [42], and the DFT-D3 method is employed to include vdW interactions [43]. Spin-orbit coupling (SOC) is included in the total-energy calculations. The FE transition pathway and energy barrier are calculated based on the nudged elastic band (NEB) approach [44]. The polarization is calculated by the modern polarization theory [45,46], and it hardly changes with the numbers of layers and calculation methods. [41] (see Section I of supplementary information for more simulation details).

The exchange coupling $J$ between magnetic moments can be evaluated by the perturbation theory [47-50]. The basic form is

$$J = \frac{<S_i|H_{ij}|S_j>}{E_j - E_i}, \qquad (1)$$

where $S_i$ and $S_j$ represent the spin states $I$ and $j$, respectively, $E_i$ and $E_j$ denote the corresponding energy levels. $H_{ij}$ represents the hopping Hamiltonian. From this equation, we can elucidate that the exchange coupling is determined by the hopping integral in the numerator, *i.e.*, the spin wavefunction overlap, and the hopping energy in the denominator, *i.e.*, the energy difference between states. In traditional bulk type-I multiferroic structures, which are periodic, electronic structures are degenerate under opposite FE polarizations, indicating neither the hopping integral nor hopping energy is changed. As a result, the ME coupling is usually weak. However, 2D vdW heterostructures made by different layered materials break such a degeneracy along the out-of-plane direction. Particularly, the exceedingly thin nature and subsequent proximity effect around interface may amplify electric-polarization impacts, *e.g.*, the switch of FE polarization can significantly tune the interlayer band alignment and denominator in Eq (1). Meanwhile, the overlap between interlayer wavefunctions may also be modified subsequently, influencing the numerator in Eq (1). If these two factors can work synergistically, the overall interlayer magnetic order can be switched, giving hope to enhanced ME coupling.

Building upon this concept, we have taken the initiative to practically realize it through first-principles calculations. In the following, we propose a 2D vdW heterobilayer consisting of monolayer CrCl$_3$ and CCPS, as depicted in Figs. 1(a) and 1(b), respectively, to achieve enhanced ME coupling. Monolayer CrCl$_3$, a member of chromium trihalides [51], manifests as a vdW insulating magnet. Each Cr ion holds a 3-$\mu_B$ in-plane magnetic moment. The intralayer magnetic order is FM, while the interlayer order is observed to be AFM [52]. The other layer of this heterostructure is CCPS, which is a member of extensively studied metal phosphorus trichalcogenides [53]. Fig. 1(b) illustrates its crystal structure, featuring a 3-$\mu_B$ in-plane moment on each Cr ion. Notably, CCPS exhibits an out-of-plane FE polarization due to off-center Cu ions, categorizing itself as a type-I multiferroic material. Both CCPS and CrCl$_3$ demonstrate insulating behaviors, characterized by band gaps of 2.4 eV and 3.7 eV, respectively [53,54]. Recent experiments have successfully obtained their monolayer structures, unveiling intriguing magneto-optical and FE properties [55,56].

Thanks to the minor lattice mismatch (0.2% ~ 0.7%) [51,53], we construct the CrCl$_3$/CCPS heterostructure in Fig. 1(c). The HSE simulation reveals that this multiferroic heterostructure exhibits both FE and magnetic orders simultaneously. Under a downward electric polarization of CCPS, the ground-state interlayer magnetic coupling is in-plane AFM. Because of the net magnetic moment of a unit cell of CrCl$_3$ is 6 $\mu_B$ (from two Cr ions) and that of CCPS is 3 $\mu_B$, the net magnetization is not zero but 3 $\mu_B$. The energy difference between the interlayer FM and AFM orders is approximately 0.25 meV per unit cell. This value is comparable to that of high-temperature (HT)-phase bilayer CrI$_3$ (~ 0.5 meV), which results in the observed interlayer AFM order [57,58]. Thus, this AFM order can be observed in similar experiments.

Upon switching the out-of-plane polarization of CCPS to the upward direction, we find that the ground-state interlayer magnetic coupling transitions to FM, accompanied by an energy gain of 0.24 meV/unit cell. The electric polarization can tune the relative magnetic energies by about 0.5 meV, which indicates the significant change for magnetic exchange interactions. Further exploration of the transition path in this heterostructure using the nudged elastic band (NEB) method, [44] as depicted in Fig. 1(d), yields an energy barrier of approximately 48 meV, aligning well with those typical values observed in FE materials [59] without the meta-stable paraelectric state [60]. Finally, the HSE band-structure calculations of those intermediate and final structures along the NEB path [41] affirm the preservation of a finite band gap (larger than 1.6 eV), a critical aspect supporting the insulating ME coupling. (see Section II of supplementary information) This feature is pivotal for experimental implementation, enabling the application of a gate electric field while minimizing the harmful leakage current.

It is known that the interlayer magnetic order is sensitive to the interlayer stacking style. There are six high-symmetry stackings styles in the CrCl$_3$/CCPS heterostructure. AA, AB$_1$, and AB$_2$ are originated from the rhombohedral phase, which is usually called the low-temperature (LT) phase in bulk chromium trihalides. AA′, AB$_1$′, and AB$_2$′ are originated from the monoclinic phase, which is known as the HT phase [41]. Our above calculation is based on the AB$_1$ stacking. We have checked the ME effect of all other interlayer configurations and summarize the result in Section III of supplementary information [41]. All stacking styles exhibit ME coupling except the LT AB$_2$-stacking style. In practice, experimental annealing and stacking procedures significantly influence the fabricated interlayer configurations. Most fabricated van der Waals (vdW) magnets are observed in the high-temperature (HT) phase [57,58,61], which encompasses the AA′, AB$_1$′, and AB$_2$′ stackings. The energy differences among these stackings are typically small, approximately 3 or 1 meV. Thus, our predicted ME effect is expected robust in fabricated CrCl$_3$/CCPS heterostructures.

Finally, we estimate the ME coupling to be about 1.25 x 10$^{-10}$ s/m. This value is comparable with those of traditional ME materials, such as MgO/Fe/Au (9.4 x 10$^{-11}$ s/m), P(VDF-TrFE)/Co/Pt (4 x 10$^{-10}$ s/m), and BiFeO$_3$ (2.1 x 10$^{-9}$ s/m) [62-64]. Moreover, the curie temperatures of CrCl$_3$ and CCPS are 17 K and 24 K, respectively, which are comparable with other two-dimensional materials, such as CrBr$_3$ (27 K) [65], Cr$_2$Ge$_2$Te$_6$ (30 K) [66], and MnBi$_2$Te$_4$ (25 K) [67].

In the following, we present the mechanisms inducing this enhanced ME coupling. Eq (1) shows that the magnetic coupling depends on both hopping integral and hopping energy. To analyze them, a clear understanding of the exchange interaction (hopping) paths is crucial, and we still choose the AB$_1$ stacking as an example to keep consistence with Fig. 1. The mechanisms are similar for other stacking styles.

Previous studies have demonstrated that the interlayer exchange interactions of vdW materials are mediated by interfacial non-magnetic atoms, so-called super-super-exchange interaction [68]. For both CrCl$_3$ and CCPS materials, due to the $C_{3v}$ symmetry under the octahedral crystal field, their five degenerate $d$ orbitals of Cr ions split into two subsets (i.e., two-fold e$_g$ levels with a higher energy and three-fold t$_{2g}$ levels with a lower energy). In the case of Cr ions with a 3-$\mu_B$ magnetic moment, all $t_{2g}$ orbitals are half-occupied, and two super-super-exchange paths can be identified [69]. Taking the $AB_1$ stacking as an example, the schematics of two interlayer super-super-exchange path are shown in Figs. 2(a) and 2(c). In Fig. 2(a), the Path-I coupling involves a hopping between two half-occupied $t_{2g}$ orbitals through the participation of two π bindings of between Cr ions and interfacial nonmagnetic atoms. Importantly, the Cr-S-Cl path forms an angle of 112 degrees, close to 90 degrees. As discovered in previously works [39,69], this nearly orthogonal $t_1$ and $t_\pi$ hoppings involve opposite spins of interfacial sulfur atoms, resulting in an AFM coupling. The corresponding schematic hopping path in the CrCl$_3$/CCPS heterostructure is presented in Fig. 2(b). In contrast, Fig. 2(c) illustrates the other interlayer interaction, Path-II. It is between a half-occupied $t_{2g}$ orbital and an empty $e_g$ orbital through the involvement of one π binding and one σ binding of Cr ions with nonmagnetic interfacial atoms, which leads to a FM coupling due to the local Hund rules [48,70]. Different from Path-I, the Cr-S-Cl hoping in Path-II is 167 degrees, close to 180 degrees. Importantly, because $t_\sigma$ and $t_2$ hoppings are nearly along the same direction, they are mediated by the same spin of the *p* orbital of interfacial sulfur atoms, as plotted in Fig. 2(d). The observed interlayer magnetic order emerges from the competition between these two distinct interlayer orbital exchange paths. There is also a third path shown in Section VI of supplementary information. [41] It has the same spin and orbital hoppings as Path-II, contributing to interlayer AFM coupling.

The reversal of FE polarization substantially impacts above hopping paths. As plotted in the left panel of Fig. 3(a), when electric polarization is downwards, both spin components are significant around interfacial sulfur atoms. In this scenario, Path-I exchange interaction becomes dominant due to the small energy difference between half-occupied $t_{2g}$ orbitals of two components, and the system favors AFM coupling. Conversely, when electric polarization is switched upwards, the left panel of Fig. 3(b) shows that there is mainly the spin-down wavefunction (blue color) around interfacial sulfur atoms while the spin-up component is nearly depleted (right panel of Fig. 3 (b)). This is elucidated in Fig. 3(c): for upward polarization, the interlayer $t_1$ hopping of Path-I is substantially quenched due to depletion of the spin-up component, reducing AFM coupling. In contrast, for the FM hopping of Path-II shown in Fig. 3 (d), the interlayer $t_2$ hopping is almost not affected because it is mediated by the spin-down component of interfacial sulfur atoms. Therefore, the overall competition favors FM coupling under upward polarization. This spin local-field effect is essentially a surface effect enhanced by the huge interface/volume ratio of vdW heterostructures, a feature unattainable in traditional bulk structures.

The other factor influencing the ME effect is the hopping energy, which is the denominator of Eq (1). In 2D out-of-plane FE materials, a significant proximity electric field is generated, and the switch of FE polarization changes the interlayer band alignment. We present the projected density of states in Fig. 4. First, we focus on the AFM Path-I coupling. Under the downward FE polarization shown in Fig. 4(a), the heterostructure exhibit a type-I band alignment with band gap of 2.1 eV. The AFM exchange Path-I is marked between occupied Cr-d-$t_{2g}$ orbitals of CCPS and $CrCl_3$. After switching FE polarization to upward (Fig. 4(b)), the heterostructure is altered to be a type-II band alignment with a band gap of 1.6 eV. It is important to see that the energy difference between those occupied Cr-d-$t_{2g}$ orbitals of CCPS and $CrCl_3$ is increased in Fig. 2(b), effectively reducing the strength of AFM Path-I coupling. In contrast, for the FM Path-II coupling marked in Figs. 4(c) and 4(d), the energy difference (~ band gap) is reduced from 2.1 eV to 1.6 eV when switching FE polarization, effectively enhancing FM coupling. As a competition result, the overall ground-state magnetic coupling tends to switch from AFM to FM, which agrees with the first-principles result. It is worth noting that these two factors, the local-field induced changes of hopping integral and band-alignment associated hopping energies work in a synergistic way. For example, the upward polarization favors the FM hopping because of both the reduced hopping energy and its almost unaffected exchange FM Path-II.

As presented in the introduction, multiferroic materials are typically categorized to type-I or type-II. In this vdW heterostructure, the coexistence of intralayer FM order and out-of-plane FE polarization within CCPS is a typical type-I multiferroic because the FE order is from off-center Cu ions while the magnetism is from Cr ions. On the other hand, the interlayer magnetic order is decided by multiple factors, such as the atomic structure, chemical environment around the surficial atoms, and electric polarization. Thus, it exhibits characteristics more akin to a type-II multiferroic, consistent with the enhanced ME coupling. However, it doesn't belong to a typical type-II multiferroic structures, such as $NiI_2$ [17,71,72]. More accurately, it exhibits that the ferroelectricity simply selects the ground-state interlayer magnetic order. Moreover, unlike the traditional type-II multiferroic materials, which usually exhibit very weak electric polarization, both magnetic order and polarization of the $CrCl_3$/CCPS heterostructure are significant. For example, the out-of-plane electric polarization is about 0.8 μC $cm^{-2}$, which is about two orders of magnitude larger than those of typical type-II multiferroic materials [17,71,72]. In this sense, the whole vdW heterostructure can be regarded as a hybrid multiferroic material.

It is worth mentioning that this hybrid multiferroics is a general phenomenon in vdW heterostructures because of the coexistence of interlayer and intralayer orders and huge interface/volume ratio, such as in $CrBr_3/CuCrP_2S_6$ and $CrCl_3/CuVP_2S_6$ heterostructure in Table S6. However, the enhanced ME coupling, which is strong enough to switch the ground-state magnetic order by FE polarization, is specific and rare. Its occurrence depends on the detailed competition of energy differences and the strength of interlayer exchange couplings of the materials involved. (see section VII of supplementary information) For material search, there are a few general principles that aid in achieving enhanced ME coupling. First, a small energy difference (e.g., less than 0.5 meV/unit cell) between FM and AFM orders is advantageous since it falls within the range of polarization-induced energy variations. Second, a highly tunable band alignment is preferred because it can be effectively modulated by out-of-plane polarization, changing the denominator term of exchange interactions.

Finally, we go beyond the bilayer heterostructure and study multilayer $CrCl_3$/CCPS heterostructures, which are relevant for experimental interests. We focus on the layer dependence of the $CrCl_3$ part. The first-principles HSE results are summarized in Fig. 5(a). Regardless of the layer number of $CrCl_3$, the ME effect is robust; the upward polarization favors a FM coupling, and the downward polarization favors an AFM coupling. This is quantitatively supported by Fig. 5(b). When the layer number of $CrCl_3$ increases, this difference gradually decreases, illustrating the ME effect is induced by the interfacial/proximity characteristics of ultra-thin vdW heterostructures.

Conversely, a macroscopic layer-dependent ME effect is observed in multilayer structures. As depicted in Fig. 5(a), for those heterostructures made by an odd layer number of $CrCl_3$ with an uncompensated AFM state, the overall magnetic moment per unit cell switches between 3 $\mu_B$ and 9 $\mu_B$ with the reversal of the FE polarization. This is the same as the heterobilayer result. However, for those heterostructures made by an even number of $CrCl_3$ layers with a compensated AFM state, switching the FE order does not alter the total magnetic moment. This is because the interlayer magnetic order between $CrCl_3$ layers remains AFM although the microscopic magnetic moment direction (Néel vector) of $CrCl_3$ layers is switched.

**Acknowledge:**

X.X. is supported by National Science Foundation (NSF) Designing Materials to Revolutionize and Engineer our Future (DMREF) DMR-2118779. L.Y. is supported by NSF DMR-2124934. The simulation used Anvil at Purdue University through allocation DMR100005 from the Advanced Cyberinfrastructure Coordination Ecosystem: Services & Support (ACCESS) program, which is supported by National Science Foundation grants #2138259, #2138286, #2138307, #2137603, and #2138296.

**Figures:**

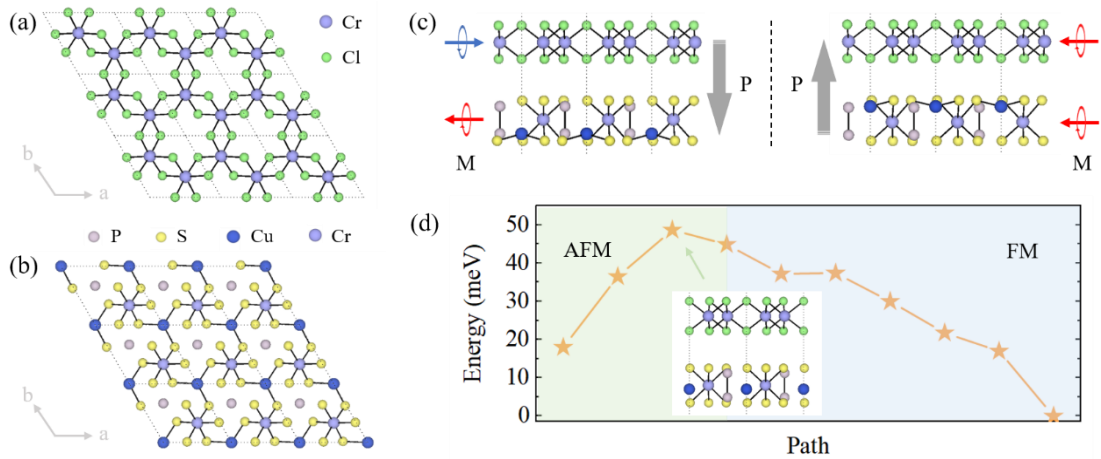

Fig. 1. (a) and (b) Top views of crystal structure of monolayer $CrCl_3$ and CCPS, respectively. (c) Side views of $CrCl_3$/CCPS heterobilayer under different electric polarization. Large arrows represent the electric polarization direction, while small arrows represent the magnetic configuration. (d) FE transition pathway between polarization-down (AFM) and polarization-up (FM). The background color illustrates the ground-state interlayer magnetic order in those transition structures.

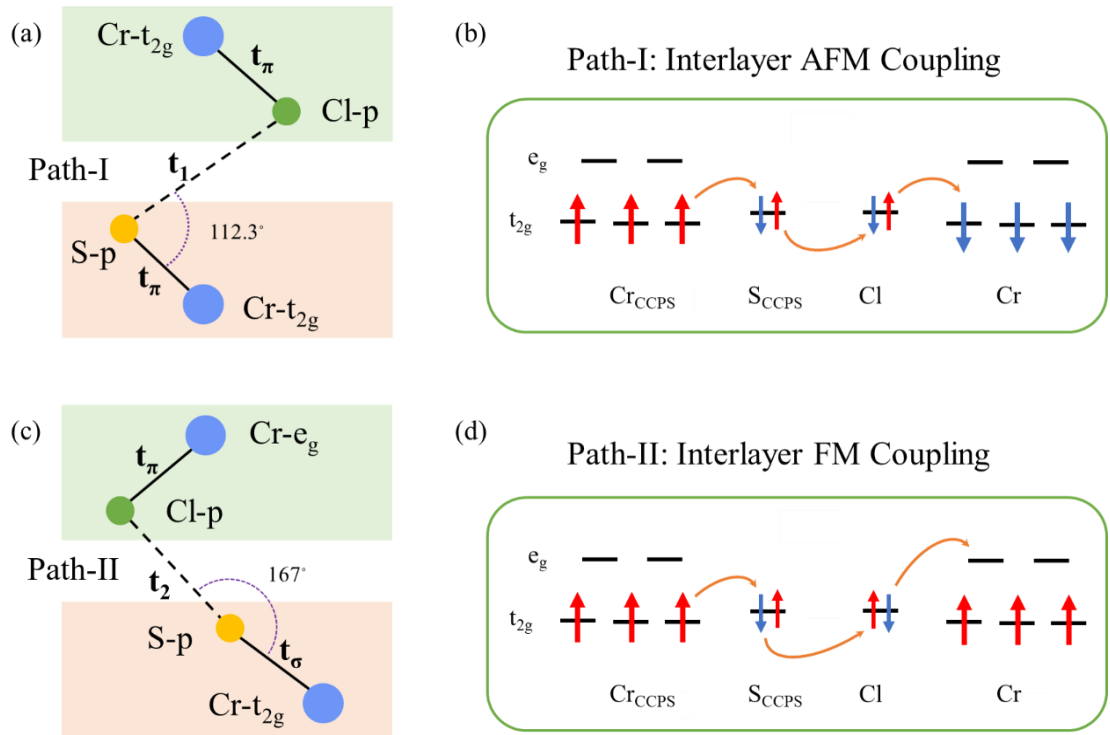

**Fig. 2.** (a) Schematic of interlayer super-super-exchange AFM path I. (b) Corresponding interlayer hoppings between spin states of path I. (c) Schematic of interlayer super-super-exchange FM path II. (d) Corresponding interlayer hoppings between spin states of path II.

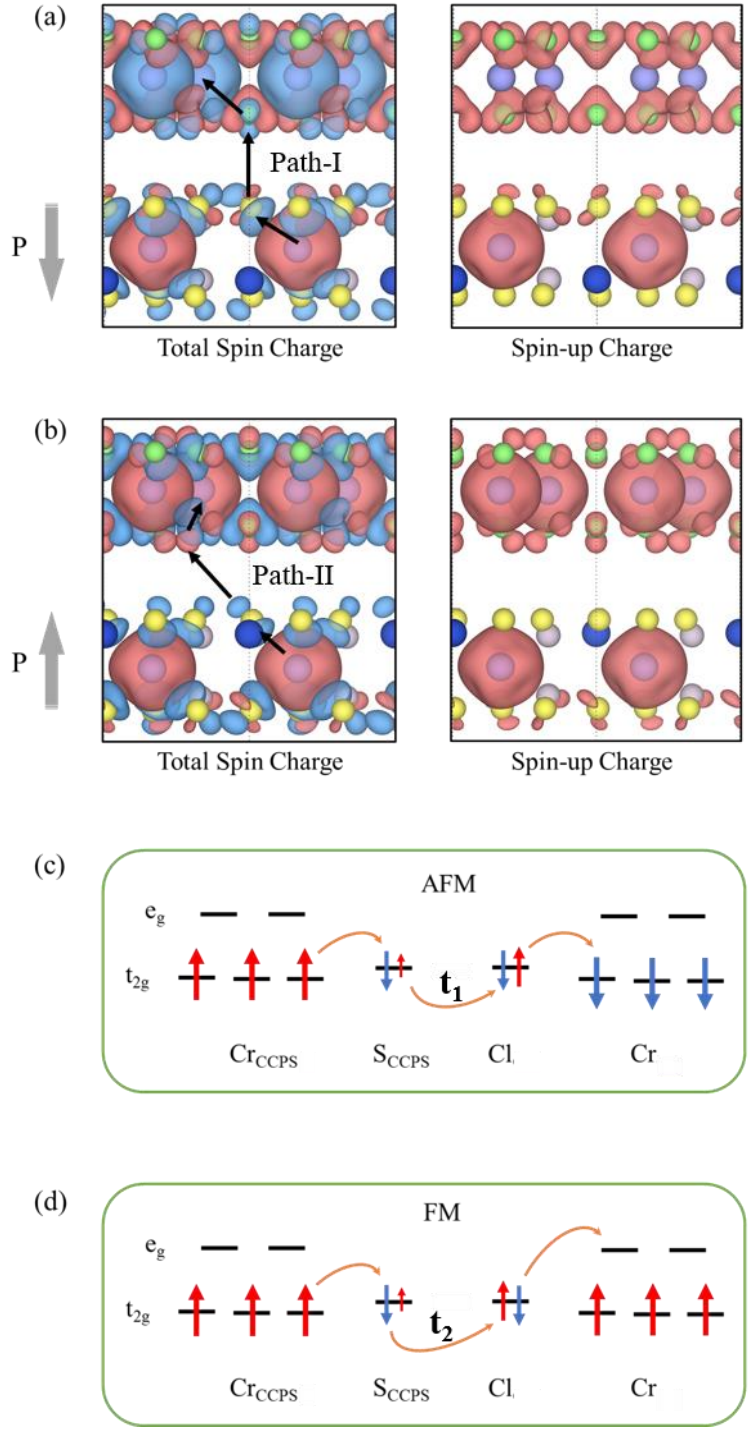

**Fig. 3.** (a) Left panel is the isosurface plot of total spin density of AFM-ground-state $CrCl_3$/CCPS under downward polarization. Right panel is the spin-up component. (b) Left panel is that of FM-ground-state $CrCl_3$/CCPS under upward polarization. Right panel is the spin-up component. The interlayer super-super-exchange paths are marked in left panels. The red and blue colors represent spin-up and spin-down components, respectively. The isosurface value is set to be 0.5% of the maximum value. (c) and (d) Schematics of AFM (path-I) and FM (path-II) interlayer hopping paths, respectively.

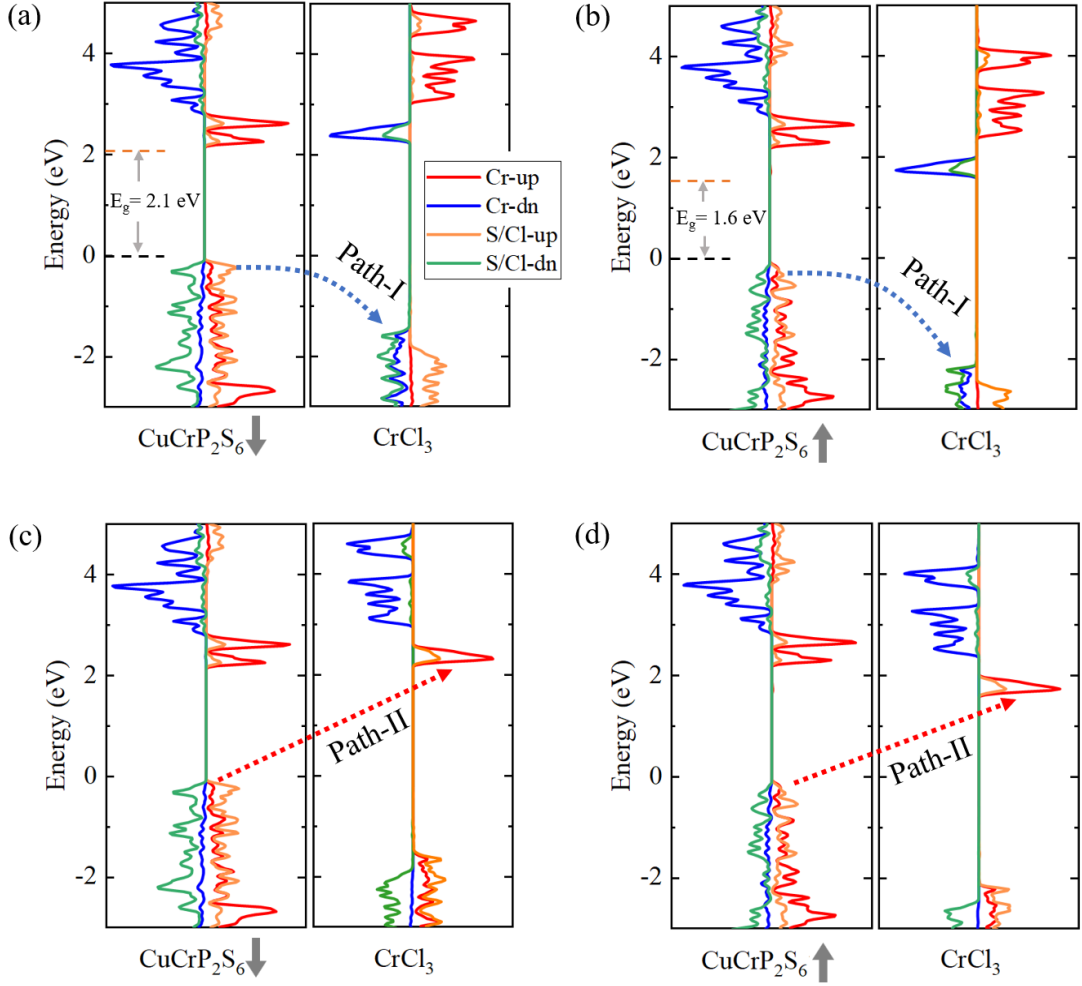

**Fig. 4.** (a) and (b) PDOS without SOC of AFM CrCl$_3$/CCPS heterostructures under downward and upward polarizations, respectively. (c) and (d) Those of AFM heterostructures under downward and upward polarization, respectively. The energy of the valence band maximum is set to be zero. The band gap and hopping paths are marked to direct readers' eyes.

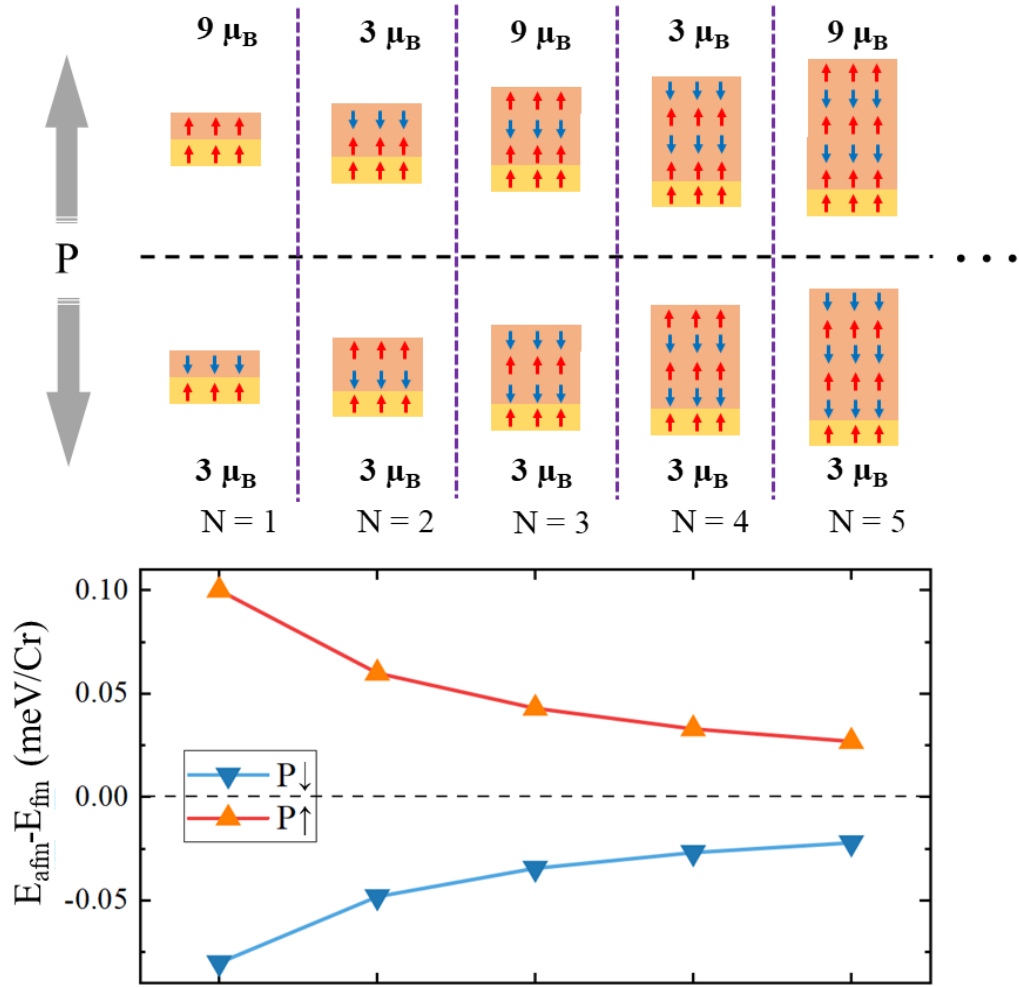

**Fig. 5**. (a) Ground-state interlayer magnetic configurations according to the CrCl$_3$ layer number (N) under opposite polarizations of CCPS. The bottom layer is fixed to be CCPS. The net magnetic moment per unit cell is listed for each configuration. (b) Corresponding total-energy differences between AFM and FM couplings according to the CrCl$_3$ layer number under opposite polarizations.